\documentclass[pra,aps,twocolumn,superscriptaddress,longbibliography,nofootinbib]{revtex4-1}

\usepackage{bm,dcolumn,amsmath,graphicx,amsfonts,amssymb}

\usepackage{xcolor}
\usepackage{float}
\usepackage[utf8]{inputenc}
\usepackage{CJKutf8}

\newcommand{\normprod}[1]{\ensuremath{:\! #1\!:}}

\newcommand{\bra}[1]{\ensuremath{\langle #1|}}	
\newcommand{\ket}[1]{\ensuremath{|#1\rangle}}	

\newcommand{\threej}[6]{\ensuremath{\begin{pmatrix}#1&#2&#3\\#4&#5&#6\end{pmatrix}}}	
\newcommand{\sixj}[6]{\ensuremath{\begin{Bmatrix}#1&#2&#3\\#4&#5&#6\end{Bmatrix}}}	

\newcommand{\boldgreek}[1]{{\mbox{\boldmath$ {#1} $}}}

\begin{document}\raggedbottom
\begin{CJK*}{UTF8}{gkai}
\title{Hyperfine structure of $^{173}\mathrm{Yb}^+$: toward resolving the $^{173}\mathrm{Yb}$ nuclear octupole moment puzzle}
\author{Di  Xiao}
	\affiliation{Department of Physics, University of Nevada, Reno, 89557, USA}
\author{Jiguang Li (李冀光)}
    \affiliation{Institute of Applied Physics and Computational Mathematics, Beijing 100088, China}
\author{Wesley C. Campbell}
    \affiliation{Department of Physics and Astronomy and Center for Quantum Science and Engineering, University of California Los Angeles, USA} 
\author{Thomas Dellaert}
    \affiliation{Department of Physics and Astronomy and Center for Quantum Science and Engineering, University of California Los Angeles, USA} 
\author{Patrick McMillin}
    \affiliation{Department of Physics and Astronomy and Center for Quantum Science and Engineering, University of California Los Angeles, USA} 
\author{Anthony Ransford}
    \affiliation{Department of Physics and Astronomy and Center for Quantum Science and Engineering, University of California Los Angeles, USA} 
\author{Conrad Roman}
    \affiliation{Department of Physics and Astronomy and Center for Quantum Science and Engineering, University of California Los Angeles, USA} 
\author{Andrei  Derevianko}
\email[]{andrei@unr.edu}
	\affiliation{Department of Physics, University of Nevada, Reno, 89557, USA}
\date{ \today }  

\begin{abstract}
Hyperfine structure (HFS) of atomic energy levels arises due 
to interactions of atomic electrons with a hierarchy of nuclear multipole moments, including magnetic dipole, electric quadrupole and higher rank moments.
Recently, a determination of the magnetic octupole moment of the $^{173}\mathrm{Yb}$ nucleus was reported from HFS measurements in neutral ${}^{173}\mathrm{Yb}$ [PRA 87, 012512 (2013)], and is four orders of magnitude larger than the nuclear theory prediction. 
Considering this substantial discrepancy between the spectroscopically  extracted value and nuclear theory, here we propose to use an alternative system to resolve this tension, a singly charged ion of the same $^{173}\mathrm{Yb}$ isotope. 
Utilizing the substantial suite of tools developed around $\mathrm{Yb}^+$ for quantum information applications, we propose to extract nuclear octupole and hexadecapole moments from measuring hyperfine  splittings in the extremely long lived first excited state ($4f^{13}(^2\!F^{o})6s^2$, $J=7/2$) of $^{173}\mathrm{Yb}^+$. 
We present results of atomic structure calculations in support of the proposed measurements. 
\end{abstract}
\maketitle
\end{CJK*}

\section{Introduction}
  \label{Sec:Intro}
While the size of an atomic nucleus is far too small to image its features directly with a microscope, the interaction of an atomic nucleus with electrons bound to it will leave signatures of the size and shape of the nucleus on the the resulting atom in the form of hyperfine structure (HFS). In particular, $P$ and $T$ symmetries dictate that the distribution of protons leads to even-rank $(k\!=\!0,2,4...)$ electric $2^k$-pole moments (e.g., monopole, quadrupole, and hexadecapole) and the distribution of currents and magnetic moments leads to odd-rank $(k\!=\!1,3,5...)$ magnetic moments (e.g.\ dipole, octupole, and 32-pole) that interact with the electrons to shift their energies.  In this sense, when combined with accurate atomic structure calculations, a measurement of the HFS of an atom constitutes an electron scattering experiment on the nucleus that allows us to ``see'' the distribution of its nucleons by observing how these well-characterized electrons scatter from it.


In general, the dominant contributions to HFS come from (nuclear) magnetic dipole and electric quadrupole interactions. Presently, the nuclear magnetic dipole ($\mu$) and electric quadrupole ($Q$) moments of most nuclei are well established (see e.g., compilation~\cite{Stone2005}). This is largely because the HFS signatures of higher-order moments only appear on electronic states with sufficiently high multiplicity $(2J \ge k)$ and the magnitude of the energy shift tends to decrease with increasing rank $k$.  The measurement of HFS signatures of high rank $(k \ge 3)$ multipoles, therefore, requires a well-controlled atom in a high angular momentum state for precision and state of the art atomic structure theory for accuracy.

Here, we focus on the potential for measuring the rarely observed nuclear octupole ($\Omega, k\!=\!3$) and hexadecapole ($\Pi, k\!=\!4$) moments. These moments have been deduced for only a handful of nuclei  and, in most cases, are in tension with nuclear theory (see Table~\ref{table:NuclearMoments}). For example, in $^{133}$Cs, the extracted~\cite{GerDerTan03} nuclear octupole moment is 40 times larger that the nuclear theory value. This paper is motivated by the even more substantial disagreement for $^{173}\mathrm{Yb}$. Recently, \citet{Singh} reported a measurement of the nuclear octupole moment from their measurements of HFS in the ${^3P_2}$ state of neutral  $^{173}\mathrm{Yb}$. However, this value $\Omega = -34.4 \,{\mathrm{b{\mu_N}}}$ is $10^4$ times larger than the nuclear theory prediction, $\Omega =0.003\,{\mathrm{b{\mu_N}}}$~\cite{Williams1962}.
This striking four orders of magnitude disagreement calls for an independent measurement and analysis. Here, we investigate the prospects for extracting $\Omega$ and higher rank nuclear multipole moments of ytterbium-173 by a combined theoretical and experimental investigation of the hyperfine level splittings in the  first excited state ($4f^{13}(^2\!F^{o})6s^2$, $J=7/2$) of the $^{173}\mathrm{Yb}^{+}$ ion. This ${}^2\!F^o_{7/2}$ state is metastable, contains six $m_F=0$ states that will be first-order insensitive to magnetic fields, and easily state-selectively coupled to the ground state for precision spectroscopy. 


\begin{table}[ht!]
\caption{
Compilation of spectroscopic determinations of nuclear octupole moments.
$\Omega^\mathrm{emp}$ are the empirical moments derived from the combination of spectroscopic HFS measurements and electronic structure calculations. 
$\Omega^\mathrm{SP}$ are octupole moments predicted by 
the single-particle model~\cite{sch55}. All octupole moments are in units  $\mathrm{barn}\times \text{nuclear magneton} (\mathrm{b\times{\mu_N}})$. 
All listed isotopes are stable except for  $^{115}\mathrm{In}$ and $^{87}\mathrm{Rb}$ with half-lives of $4.4\times10^{14}$ and $4\times10^{10}$ years, respectively~\cite{Stone2005}.
Values of $\Omega$ have also been reported for about 20 additional nuclei from nuclear scattering experiments~\cite{Fuller1976}. 
}
\label{table:NuclearMoments}
 \begin{tabular}{ccccrr}
  \hline
  \hline
 \multicolumn{1}{c}{Isotope}&
 \multicolumn{1}{c}{$\mathrm{I^{\pi}}$}&
 \multicolumn{1}{c}{Valence nucleon}&
 \multicolumn{1}{c}{Atomic state}&
 \multicolumn{1}{c}{$\mathrm{\Omega^{emp}}$} &
 \multicolumn{1}{c}{$\mathrm{\Omega^{SP}}$} \\
\hline
$^{87}\mathrm{Rb}$~\cite{Gerginov2009}&$\frac{3}{2}^{-}$ &$p_{3/2}, \mathrm{proton}$ &  $^2\!P_{{3}/{2}}$& $-0.58$ & $0.30$\\
  $^{113}\mathrm{In}$~\cite{EckKus57} &$\frac{9}{2}^{+}$  &
  $g_{9/2}, \mathrm{proton}$ & $^2\!P_{3/2}$&$0.574$ &$0.99$ \\
 $^{115}\mathrm{In}$~\cite{EckKus57} & $\frac{9}{2}^{+}$ & 
 $g_{9/2}, \mathrm{proton}$&$^2\!P_{{3/2}}$ &$0.565$ &$1.00$ \\
 $^{133}\mathrm{Cs}$~\cite{GerDerTan03} & $\frac{7}{2}^{+}$  & $g_{7/2}, \mathrm{proton}$ & $^{2}\!P_{{3/2}}$ & $0.82$ &$0.022$ \\
 $^{137}\mathrm{Ba^{+}}$~\cite{Lewty2013} &$\frac{3}{2}^{+}$ &$d_{3/2}, \mathrm{neutron}$& $^{2}\!D_{{3/2}}$ & $-0.0629$ &$0.039$ \\
 $^{155}\mathrm{Gd}$~\cite{Unsworth1969} & $\frac{3}{2}^{-}$  & $p_{3/2}, \mathrm{neutron}$ & $^9\!D_3$ & $-1.66$ & $-0.29$ \\
 $^{165}\mathrm{Ho}$~\cite{DanFerGeb74}& $\frac{7}{2}^{-}$& $f_{7/2}, \mathrm{proton}$& $^4\!I_{{15/2}}$&$0.75$ &$1.0$\\
 $^{173}\mathrm{Yb}$~\cite{Singh} &$\frac{5}{2}^{-}$ &$f_{5/2}, \mathrm{neutron}$ &$^{3}\!P_2$ & $-34.4$ & $0.15$ \\
\hline
\hline
\end{tabular}
\end{table}


Most of the previous spectroscopic determinations of high-order moments of nuclei focused on extraction of octupole moments, and are compiled in Table \ref{table:NuclearMoments}. This Table also lists the nuclear single-particle  model~\cite{sch55} values for the  nuclear octupole moments. In addition to the listed spectroscopic determinations, the experiments were
carried  out in Eu~\cite{Chi91} and Hf~\cite{JinWakIna95}. However,
due to the complexity of electronic structure calculations, these experiments only determined ratios of nuclear octupole moments between different isotopes, $\Omega(^{151}\mathrm{Eu})/\Omega(^{153}\mathrm{Eu})$ and $\Omega(^{177}\mathrm{Hf})/\Omega(^{179}\mathrm{Hf})$.

Beyond octupole order, the hexadecapole moment $\Pi$ has been spectroscopically determined for only one species: $^{165}$Ho~\cite{DanFerGeb74}.
Access to the hexadecapole moments requires $J \ge 2$ and $I \ge 2$.
For example, although $^{133}$Cs nucleus has $I=7/2$ and thereby possesses hexadecapole moment, this moment can not be determined from the measured HFS of the $6\,{p_{3/2}}$ state~\cite{GerDerTan03}. This argument prohibits the extraction of nuclear hexadecapole moments from the structure of the states used to measure magnetic octupole moments for all but two exceptions in Table~\ref{table:NuclearMoments}: 
$^{165}$Ho and $^{173}$Yb. The value for $\Pi$ that was extracted from spectroscopic measurements in 
$^{165}$Ho was found to be larger than the nuclear theory value by an order of magnitude~\cite{DanFerGeb74}. In principle, one could extract $\Pi$ from the measurements made by Singh \textit{et al.} in neutral $^{173}\mathrm{Yb}$ \cite{Singh}, but its contribution was neglected in that work. Here, in order to leverage the considerable experimental toolbox built around $\mathrm{Yb}^+$ for quantum information applications, we evaluate the necessary electronic structure factors for $^{173}$Yb$^{+}$ needed to enable extraction of the hexadecapole moment of this isotope from future spectroscopic measurements.

The $\mathrm{Yb}^{+}$ ground state hyperfine structure is among the most precisely measured and easiest to control of all the HFS in atomic physics owing to its $m_F=0$ ``clock states'' and its readily available state preparation and readout schemes. The ${}^2S_{1/2}$ HFS of ${}^{171}\mathrm{Yb}^+$ has been used for decades for frequency standards and quantum information processors, and its splitting has been known to $\mathrm{mHz}$ precision for many years \cite{Fisk1997}. The long coherence time of clock-state qubits defined on this hyperfine splitting (recently shown to exceed 10 minutes \cite{Wang2017}) has made ${}^{171}\mathrm{Yb}^{+}$ a premier qubit host for quantum computing and quantum simulation \cite{Olmschenk2007, Figgatt2019, Zhang2017, Landsman2019, Wright2019}. Likewise, the metastable $^2F_{7/2}^o$ electronic state of ${}^{171}\mathrm{Yb}^+$ lives for years, and the E3 transition on ${}^2S_{1/2} \! \leftrightarrow {}^2F_{7/2}^o$ is used as an optical frequency standard \cite{Huntemann2016}, where hyperfine structure within these states allows control of systematics. Some of the current best limits on the time variation of fundamental constants are based on precision measurements between specific hyperfine components of this E3 transition \cite{Godun2014, Huntemann2014b}.  


As the experimental progress with this species continues to achieve higher accuracy and precision \cite{Huntemann2016, Sanner2018}, theoretical work is needed in parallel with these improvements to understand contributions to systematics. Ref.~\cite{Beloy2008a} has shown how to calculate the second-order energy correction due to hyperfine interaction for the alkali-atoms in the first excited state, and gives the theoretical basis for higher-order terms calculated in this paper. The values of hyperfine constants $A$ and $B$ for $^{171}\mathrm{Yb}^{+}$ and $^{173}\mathrm{Yb}^{+}$ in the ${}^2F_{7/2}^o$ state are also given in Ref.~\cite{Dzuba2016a}, and are used as a comparison to our values.

%

The paper is organized as follows. In Sec.~\ref{Sec:General-theory}, we review the theory of hyperfine structure. Based on this general theory, we derive the first- and second-order corrections to the HFS of $^{173}\mathrm{Yb^{+}}$ in the first excited state. In Sec.~\ref{Sec:electronic-reduced-ml}, we compute $^{173}\mathrm{Yb^{+}}$ electronic-structure factors required for extracting nuclear moments. We discuss the importance of correlation effects in Sec.~\ref{Sec:electron-correlation-eff}. 
Finally, we estimate theoretical accuracy and consider its implications on the extraction of octupole and hexadecapole moments in Sec.~\ref{sec:Discussion}.
%
Unless specified otherwise, atomic units are used throughout.

\section{Review of the theory of hyperfine structure}
  \label{Sec:General-theory}
The hyperfine interaction can be decomposed into the magnetic dipole (M1), electric quadrupole (E2), magnetic octupole (M3), electric hexadecapole (E4), and higher rank contributions.
We start by expressing the hyperfine  Hamiltonian in irreducible tensor form~\cite{sch55,Johnson2007}
 \begin{equation}
     H_\mathrm{HFI}=\sum\limits_{k,\mu}(-1)^{\mu}T_{k,\mu}^eT_{k,-\mu}^n \,,\ \label{Eq:HF-general}
 \end{equation}
where rank-$k$ tensors $T_{k,\mu}^e$ act in the electron space, and $T^n_{k,-\mu}$ --- in the nuclear space. The many-electron operators are $T_{k,\mu}^e=\sum\limits_{i}t_{k,\mu}^e(i)$, where the summation is over all the atomic electrons. The single-electron operators $t_{k,\mu}^e(i)$ can be divided into two groups~\cite{Johnson2007},
  \begin{equation}
      t_{k,\mu}^e(i)= \begin{cases}
      -\frac{1}{r^{k+1}}C_{k,\mu}(\hat{r}), & \text{electric (even $k$)}\,, \\
      -\frac{i}{r^{k+1}}\sqrt{\frac{k+1}{k}} \bm{\alpha} \cdot{\bm{C}^{(0)}_{k,\mu}}(\hat{r}), & \text{magnetic (odd $k$)}. 
      \end{cases}\label{Eq:single-electron-operator}
  \end{equation}
Here,  $\bm{\alpha}$ is the Dirac  matrix, $r$ is the radial coordinate, $C_{k,\mu}$ are normalized spherical harmonics, and $\bm{C}_{k,\mu}^{(0)}$ are normalized vector spherical harmonics. 

 The first-order energy correction due to hyperfine interaction, Eq.~(\ref{Eq:HF-general}), in the basis of coupled nuclear and atomic states is~\cite{Beloy2008a}
\begin{equation}
\begin{split}
 W^{(1)}_F = \langle \gamma{IJFM_F}|H_{\mathrm{HFI}}|\gamma IJFM_F\rangle = (-1)^{I+J+F} \\
    \times\sum\limits_{k}\begin{Bmatrix}
    F & J & I \\
    k & I & J
    \end{Bmatrix}\langle \gamma{J}||T^e_k||\gamma{J}\rangle\langle I||T^{n}_k||I\rangle\,,
\end{split}\label{Eq:1st-order-ME}
\end{equation}
where $I$ is the nuclear spin, $J$ is the total electronic angular momentum, $F$ is the grand total angular momentum $\bm{F}=\bm{J} +\bm{I}$, and 
$\gamma$ stands for remaining quantum numbers.
 
The first-order energy corrections are conventionally expressed as linear combinations of HFS constants $A$, $B$, $C$, $D$...($k$=1,2,3,4...). The first four constants are defined as~\cite{Beloy2008a}
 \begin{eqnarray}\label{eq:hpf-constants}
     A&=&\frac{1}{IJ}\langle{T_1^n}\rangle_{I}\langle{T_1^e}\rangle_{J}=\frac{1}{IJ}\mu\langle{T_1^e}\rangle_J\,,\nonumber     \\
     B&=&4\langle{T_2^n}\rangle_I\langle{T_2^e}\rangle{_J}=2Q\langle{T_2^e}\rangle_{J}\,,\\ 
     C&=&\langle{T_3^n}\rangle_I\langle{T_3^e}\rangle{_J}=-\Omega\langle{T_3^e}\rangle{_J}\,,\nonumber\\
     D&=&\langle{T_4^n}\rangle{_I}\langle{T_4^e}\rangle{_J}=\Pi{\langle{T_4^e}}\rangle{_J} \,.\nonumber 
 \end{eqnarray}
Here, the stretched matrix element $\langle T_k^e \rangle_J$ is defined as $\langle T_k^e \rangle_J =\threej{J}{k}{J}{-J}{0}{J}\bra{\gamma{J}}|T_{k}^e|\ket{\gamma{J}}$. Nuclear stretched matrix elements are proportional to the nuclear moments:
$\langle{T_1^n}\rangle_I=\mu$,
$\langle{T_2^n}\rangle_I=Q/2$,
$\langle{T_3^n}\rangle_I=-\Omega$, and                  $\langle{T_4^n}\rangle_I=\Pi$.

The second-order energy correction due to hyperfine interaction reads
\begin{widetext}
	\begin{eqnarray}\label{eq:2nd-order-general}
	     W^{(2)}_{F}&=&\sum_{\gamma{'}J^{'}}\frac{\langle\gamma{IJFM_F}|H_{\mathrm{HFI}}|\gamma^{'}IJ'FM_F\rangle\langle\gamma^{'}IJ'FM_F|H_{\mathrm{HFI}}|\gamma{IJFM_F}\rangle}{E_{\gamma{J}}-E_{\gamma{'}J'}}.
	\end{eqnarray}
\end{widetext}
This equation reduces to
	\begin{equation}\label{eq:2nd-order}
	\begin{split}
	    W_F^{(2)} = \sum\limits_{\gamma'J'}\frac{1}{E_{\gamma{J}}-E_{\gamma{'}J'}}\sum\limits_{k_1,k2}\sixj{I}{J}{F}{J'}{I}{k_1}\sixj{I}{J}{F}{J'}{I}{k_2} \times \\
	    \bra{I} |T_{k_1}^n|\ket{I}\bra{I} |T_{k_2}^n|\ket{I}\bra{\gamma{J}}|T_{k_1}^e|\ket{\gamma'J'}\bra{\gamma J}|T_{k_2}^e| \ket{\gamma' J'},
	\end{split}
	\end{equation}
	where primed quantities refer to intermediate states; $E_{\gamma{J}}$ and $E_{\gamma{'}J{'}}$ are the HFI-unperturbed energy levels. 
	
Based on the general theory, in the next section we investigate the hyperfine structure of $^{173}\mathrm{Yb^{+}}$ in the first excited state. 

\section{Hyperfine structure of  $\mathrm{Yb}^{+}$ in the first excited state}
\label{Sec:HFI-YbII}

   The first excited state of $\mathrm{Yb^{+}}$  has the electronic configuration $4f^{13}(^2\!F^o)6s^2$ with electronic angular momentum $J$ equal to $7/2$. Since $^{173}\mathrm{Yb}$ has nuclear spin of $5/2$, the grand total angular momentum $F$ is an integer in the interval $[1,6]$. The $^{173}\mathrm{Yb}$ isotope possesses five distinct nuclear electromagnetic moments. The nucleus has an unpaired valence neutron in the $f_{5/2}$ state.
   The observed~\cite{Stone2005} nuclear magnetic dipole  $\mu$ and electric quadrupole moments $Q$ are equal to $-0.680\,\mathrm{\mu_N}$ and $2.80\mathrm{\,b^2}$, 
   respectively. The nuclear single-particle shell model is not adequate for this isotope as it predicts zero value for the quadrupole moment (the valence nucleon is a neutron for this isotope, whereas the electric moments arise from the distribution of protons in the core). This discrepancy points to a strong nuclear deformation of $^{173}\mathrm{Yb}$.
   Following the theoretical proposal~\cite{BelDerJoh08}, the value for the octupole moment was deduced~\cite{Singh} from the HFS in neutral $^{173}\mathrm{Yb}$ atom in the metastable $6s6p\,^3\!P_2$ state. 
   However, the deduced value, $\Omega = -34.4 \,{\mathrm{b\times{\mu_N}}}$, is $\sim 200$ times larger and of opposite sign compared to the prediction of the single-particle nuclear shell model~\cite{Schwartz1955}. A more sophisticated nuclear structure calculation~\cite{Williams1962} (axially-symmetric collective model in strong coupling) yields $\Omega =0.003\,{\mathrm{b\times{\mu_N}}}$, bringing the discrepancy with the spectroscopic determination in neutral Yb to four orders of magnitudes. As to the electric hexadecapole moment $\Pi$, the single-particle nuclear shell model again predicts zero (similar to $Q$) because the valence nucleon is electrically neutral. We are not aware of any nuclear structure calculations for $\Pi$ of $^{173}\mathrm{Yb}$. We estimate $\Pi \approx Q^2 \approx 9 \, \mathrm{b^2}$ as both $Q$ and $\Pi$ arise due to nuclear deformation; we will take this value as fiducial in further computations.

   
   From Eqs.~(\ref{Eq:1st-order-ME},\ref{eq:hpf-constants}), we obtain the following first-order energy corrections,
     \begin{eqnarray} 
	W_6^{(1)}&=&\frac{35}{4}A+\frac{1}{4}B+C+D	\, , \nonumber \\
	W_5^{(1)}&=&\frac{11}{4}A-\frac{37 }{140}B-\frac{109}{35}C-\frac{41}{7}D \, ,\nonumber  \\
	W_4^{(1)}&=&-\frac{9}{4}A-\frac{3}{10}B+\frac{46}{35}C+12 D\, , \nonumber \\
	W_3^{(1)}&=&-\frac{25}{4}A-\frac{1}{14}B+\frac{22}{7}C-\frac{44}{7}D \, , \label{Eq:W1-individual}\\
	W_2^{(1)}&=&-\frac{37}{4}A+\frac{1}{4}B+\frac{11}{35}C-11D \, ,\nonumber  \\
	W_1^{(1)}&=&-\frac{45}{4}A+\frac{15}{28}B-\frac{33}{7}C+\frac{99}{7}D \, \nonumber .
\end{eqnarray}

 The second-order corrections are computed from Eqs.~(\ref{eq:2nd-order-general},\ref{eq:2nd-order}), where we keep magnetic dipole and electric quadrupole contributions. To streamline the notation, 
 we introduce dipole-dipole, dipole-quadrupole, and quadrupole-quadrupole constants. These are defined for individual intermediate states $|\gamma'J'\rangle$,
 \begin{widetext}
\begin{eqnarray}\label{eq:2nd-order-term}
    \eta_{\mu\mu}[\gamma'J']&=&\frac{(I+1)(2I+1)}{I}\frac{\mu^2\bra{\gamma J} |T_1^e|\ket{\gamma' {J{'}}}^2}{E_{\gamma{J}}-E_{\gamma' J' }}\,,\nonumber \\
    \eta_{\mu{Q}}[\gamma'J']&=&{\frac{(I+1)(2I+1)}{I}}\sqrt{\frac{2I+3}{2I-1}}\times\frac{\mu{Q}\bra{\gamma{J}}|T_1^e|\ket{\gamma'{J'}}\bra{\gamma{J}}|T_2^e|\ket{\gamma'{J'}}}{E_{\gamma{J}}-E_{\gamma'{J'}}}\,, \\
    \eta_{QQ}[\gamma'J']&=&\frac{(2I+1)(I+1)(2I+3)}{4I(2I-1)}\frac{Q^2\bra{\gamma{J}}|T_2^e|\ket{\gamma' {J'}}^2}
    {E_{\gamma{J}}-E_{\gamma' {J'}}}\,.\nonumber
\end{eqnarray}
\end{widetext}
Eq.~(\ref{eq:2nd-order}) shows that we need to sum over all possible intermediate states obeying both the parity and the angular selection rules - that is, the parity of the $\ket{\gamma{J}}$ and $\ket{\gamma{'}{J'}}$ states has to be the same and $|J+J'|\geq{k}\geq|J-J'|$. Thus, for dipole-dipole and dipole-quadrupole terms, there are three possible $J'$ values, while for quadrupole-quadrupole term, there are five possible $J'$ values. Among the $\ket{\gamma{'}{J'}}$ intermediate states, the dominant contribution comes from the configuration $4f^{13}(^2\!F^{o})6s^2$ with $J=5/2$. The electronic matrix elements of other possible intermediate states are small enough to be neglected, or the energy denominators are large.
With the single intermediate state fixed, we rewrite Eq.~(\ref{eq:2nd-order}) as
\begin{eqnarray}
W_F^{(2)}&\approx &C_{\mu{\mu}}[J',F]\times\eta_{\mu{\mu}}[\gamma{'}J']+ \nonumber \\
&& C_{\mu{Q}}[J',F]\times\eta_{\mu{Q}}[\gamma'J'] + \label{Eq:W2-individual} \\
&&C_{QQ}[J',F]\times\eta_{QQ}[\gamma{'}J']\,,\nonumber
\end{eqnarray}
where the angular factors are 
\begin{eqnarray}
C_{{\mu\mu}}[J{'},F]&=&\sixj{I}{J}{F}{J'}{I}{1}^2\,,\nonumber\\
C_{{\mu{Q}}}[J{'},F]&=&\sixj{I}{J}{F}{J'}{I}{1}\sixj{I}{J}{F}{J'}{I}{2}\,,\label{eq:2nd-order-coeff}\\
C_{{{Q}{Q}}}[J{'},F]&=&\sixj{I}{J}{F}{J'}{I}{2}^2 \,.\nonumber
\end{eqnarray}
Adding the first-order, Eq.~(\ref{Eq:W1-individual}) and second-order, Eq~(\ref{Eq:W2-individual}), corrections for individual hyperfine levels, we arrive at
\begin{eqnarray}
 \label{Eq:W12-individual}
 	W_6^{(1+2)}&=&W_6^{(1)}+0\times{\eta_{\mu\mu}}+0\times{\eta_{\mu{Q}}}+0\times\eta_{QQ}\,,\nonumber\\	
	W_5^{(1+2)}&=&W_5^{(1 )}+\frac{1}{98}\eta_{\mu\mu}+\frac{\sqrt\frac{5}{6}}{98}\eta_{\mu{Q}}+\frac{5}{588}\eta_{QQ}\,,\nonumber\\
	W_4^{(1+2)}&=&W_4^{(1)}+\frac{11}{882}\eta_{\mu\mu}+0\times\eta_{\mu{Q}}+0\times\eta_{QQ}\,,\nonumber\\
	W_3^{(1+2)}&=&W_3^{(1)}+\frac{1}{98}\eta_{\mu\mu}-\frac{\sqrt{\frac{2}{15}}}{49}{\eta_{\mu{Q}}}+\frac{4}{735}\eta_{QQ}\,,\\
	W_2^{(1+2)}&=&W_2^{(1)}+\frac{3}{490}\eta_{\mu\mu}-\frac{\sqrt{\frac{3}{10}}}{70}\eta_{\mu{Q}}+\frac{1}{100}\eta_{QQ}\,,\nonumber\\
	W_1^{(1+
	2)}&=&W_1^{(1)}+\frac{1}{441}\eta_{\mu\mu}-\frac{1}{{49}\sqrt{30}}{\eta_{\mu{Q}}}+\frac{3}{490}\eta_{QQ}\,.\nonumber
 \end{eqnarray}

Experimentally relevant quantities are the HFS energy intervals $\Delta{W}_F = W_{F+1} - W_F$. Explicitly,
\begin{widetext}
\begin{eqnarray} \label{eq:DeltaW12}  
\Delta{W}_5^{(1+2)}&=&6 A+\frac{18 }{35}B+\frac{144 }{35}C+\frac{48 }{7}D-\frac{1}{98}\eta_{\mu\mu} -\frac{1}{98} \sqrt{\frac{5}{6}} \eta_{\mu{Q}}-\frac{5}{588}\eta_{{Q}{Q}} \,, \nonumber\\
\Delta{W}_4^{(1+2)}&=&5 A+\frac{1}{28}B-\frac{31 }{7}C-\frac{125 }{7}D-\frac{\eta_{\mu\mu}}{441} +\frac{1}{98} \sqrt{\frac{5}{6}} \eta_{\mu{Q}}+\frac{5}{588}\eta_{{Q}{Q}}\,,\nonumber\\
\Delta{W}_3^{(1+2)}&=&4 A-\frac{8 }{35}B-\frac{64 }{35}C+\frac{128 }{7}D+\frac{\eta_{\mu\mu}}{441}+\frac{1}{49} \sqrt{\frac{2}{15}} \eta_{\mu{Q}} -\frac{4}{735}\eta_{{Q}{Q}} \,,\\
\Delta{W}_2^{(1+2)}&=&3 A-\frac{9 }{28}B+\frac{99 }{35}C+\frac{33 }{7}D+\frac{1}{245} \eta_{\mu\mu}+(\frac{1}{70} \sqrt{\frac{3}{10}} -\frac{1}{49} \sqrt{\frac{2}{15}}) \eta_{\mu{Q}}-\frac{67}{14700}\eta_{{Q}{Q}}\,,\nonumber\\
\Delta{W}_1^{(1+2)}&=&2 A-\frac{2 }{7}B+\frac{176 }{35}C-\frac{176 }{7}D+\frac{17 }{4410}\eta_{\mu\mu} +(\frac{1}{49 \sqrt{30}}-\frac{1}{70} \sqrt{\frac{3}{10}}) \eta_{\mu{Q}}+\frac{19}{4900}\eta_{{Q}{Q}}\,.\nonumber
 \end{eqnarray}
\end{widetext}

To determine the HFS constants $A,B,C$, and $D$ from  experimental measurements of $\Delta{W}_F$, in  Sec.~\ref{Sec:electronic-reduced} we compute the second-order corrections. Further, to find the values of nuclear octupole and hexadecapole moments from $C$ and $D$  we need electronic form-factors; these are also computed in Sec.~\ref{Sec:electronic-reduced}.
We neglect contributions of one remaining HFS constant $E$ arising from the $2^5$-pole nuclear magnetic moment. This contribution is expected to be strongly suppressed compared to the contribution of the octupole moment (see Sec.~\ref{sec:Discussion}).

    

\section{Calculations of electronic structure factors} 
\label{Sec:electronic-reduced}
 \subsection{Dirac-Hartree-Fock calculations}
 \label{Sec:electronic-reduced-ml}
$\mathrm{Yb^{+}}$ ion in the first excited state contains thirteen $4f$ electrons and two $6s$ electrons. In this section, we start our calculation of the electronic wave functions by employing the frozen core Dirac-Hartree-Fock (DHF) approximation. In this approximation, we compute the DHF orbitals of the $\mathrm{YbIII}\,([\mathrm{Xe}]4f^{14})$ core. Then the valence (outside the $\mathrm{[Xe]}4f^{14}$ core) orbitals are computed using the DHF potential of the core. The many-body wave function $\psi_{J\,,M}$ can be approximated as 
\begin{eqnarray}\label{eq:multi-electron-wf}
&&\ket{\psi_{J,\,M}}\simeq \frac{1}{2} (-1)^{7/2-M}\times\nonumber\\
&&(\sum\limits_{m}(-1)^{m-1/2}{a_{6s_{1/2,m}}^{\dagger}a_{6s_{1/2,-m}}^{\dagger})\,a_{4f_{7/2,-M}}}\ket{{0_c}}\,,
\end{eqnarray}where $a_{6s_{1/2},m}^{\dagger}$ are creation operators with magnetic quantum number $m$ equal to either $-1/2$ or $1/2$, $a_{4f_{7/2,\,M}}$ is an annihilation operator for the $4f_{7/2}$ orbital, and $\ket{0_c}$ represents the $\mathrm{[Xe]}4f^{14}$ core. The phase factor $(-1)^{7/2-M}$ is generated after moving the hole operator from the core state~\cite{Johnson2007}. The two $6s_{1/2}$ orbitals are coupled so that the $6s^2$ valence shell has zero value of angular momentum. 
Using Wick's theorem,  we write the matrix element~(\ref{Eq:single-electron-operator}) in the multi-electron state as an expectation value in the hole orbital (see Appendix~{\ref{app:2nd-quantization}} for derivation) 
	\begin{equation}
	    \bra {\psi_{J,\,M}} T^e_{k,\,\mu} \ket{\psi_{J,\,M}}  = -\bra{\phi_{J,\,-M}}t_{k,\,\mu}^e \ket{\phi_{J,\,-M}},\label{eq:multi-electron-state}
	\end{equation}
where  $\ket{\phi_{J,\,-M}}$ represents the $4f_{7/2}$ hole orbital with $J$ and $-M$ being the electron's angular momentum and magnetic quantum number. The electronic tensors $T_{k,\,\mu}^e$ are given by Eq.~(\ref{Eq:HF-general}).
In Appendix~{\ref{app:2nd-quantization}}, we show that the reduced matrix elements are related as
\begin{eqnarray}
\label{eq:relation_multi_hole}
\bra{\psi_{J}}|T^e_{k,\mu}|\ket{\psi_{J}}=(-1)^{k+1}\bra{\phi_{J}}|t_k^e|\ket{\phi_{J}}\,,
\end{eqnarray}
with reduced matrix elements specified in Appendix~\ref{app:matrix-element}.
The transition from a multi-electron state to the single-electron hole orbital greatly simplifies our calculation since it only requires the one-electron $4f_{7/2}$ orbital, which can be easily obtained self-consistently with the DHF method. 
Our computed values of the first- and second-order hyperfine constants are listed in the first row of Table~\ref{Table:HFS}.
	
\begin{table*}[!ht]
\caption{\label{Table:HFS} First-order and dominant second-order hyperfine constants (in MHz) for the $4f^{13}5s^2~^2\!F_{7/2}$ state of $^{173}\mathrm{Yb}^{+}$. $C/\Omega$ is in $\mathrm{MHz/(b\times{\mu_N}})$ and $D/{\Pi}$ is in $\mathrm{MHz/b^2}$. We used the values~\cite{Stone2005} for the nuclear magnetic dipole and electric quadrupole moments, $\mu=-0.68\,\mu_{N}$  and $Q=2.80(4)~\mathrm{b}$. 
}
\begin{tabular}{cccccccc}
\hline 
\hline
Method   &    $A$  & $B$&   $C/{\Omega}$  &   $D/{\Pi}$  & $\eta_{\mu\mu}$& $\eta_{\mu{Q}}$ &$\eta_{QQ}$\\
\hline
\multicolumn{8}{c}{This work}\\ 
DHF      &  $-239$   &  $-5330$   &  ~$4.53\times 10^{-4} $  & $2.00\times{10^{-4}}$ &  $-1.50\times{10^{-2}}$ & $-0.112$  &  $-0.209$ \\      
DHF (GRASP)                         &  $-252$   &   $-5622$  &  ~$4.83 \times 10^{-4}$  & $2.25\times{10^{-4}}$ &  $-1.91\times{10^{-2}}$ &$-0.124$ &$-0.200$\\
MCDHF                      &  $-241$   &   $-5061$  &  $-6 \times 10^{-4}$  &  $2.35\times{10^{-4}}$ &                         &           & \\
\hline
\multicolumn{8}{c}{Prior work}\\ 
CI+MBPT, Ref.~\cite{Dzuba2016a} &     $-240$     &    $-4762$    &               &            & &&\\
{MCDHF}, Ref~\cite{Petrasiunas2012}  & $-304$ &$-3680$ & & & & &\\
\hline     
\hline
\end{tabular}
\end{table*}

\subsection{Electron correlation effects}
\label{Sec:electron-correlation-eff}
We employ the multi-configuration Dirac-Hartree-Fock (MCDHF) method~\cite{Grant2007, FroeseFischer2016a} to capture the main electron correlations in the Yb$^+$ ion. In this approach, an atomic state wave-function (ASF) is represented as a linear combination of configuration state functions (CSFs) with the same parity, total angular momentum, and its component along the quantization axis. The CSFs are generated by single (S) and double (D) substitutions of orbitals occupied in the reference configurations with virtual orbitals. The reference configurations constitute the dominant CSFs of the ASF concerned. The MCDHF calculation starts from the optimization on occupied orbitals in the reference configurations. By contrast to Sec~\ref{Sec:electronic-reduced-ml}, 
all of these orbitals are generated in the self-consistent field procedure. Virtual orbitals are augmented layer by layer in order to monitor the convergence of level energies and other atomic properties. Each layer includes orbitals with different angular symmetries. In addition, only the virtual orbitals in the latest added layer are variable. The details of computational strategies can be found in Ref.~\cite{Bieron2009, Li2012}.

In our calculations, 
{
we adopt the extended optimal level (EOL) scheme to optimize the two states of the [Xe]$4f^{13}6s^2$ configuration simultaneously.} The electron correlations in the $4f$ and $6s$ valence subshells and the correlations between electrons in the valence and $n=3,4$ core subshells were accounted for by CSFs 
generated by the SD replacement of the $n \ge 3$ occupied orbitals in the reference configuration with the virtual orbitals. The double replacements were 
restricted to 
only a single electron of the core subshells being promoted into the virtual orbitals at a time. The final set of virtual orbitals is composed of five  orbitals per each of the  $s, p, d, f, g, h, i$ angular momenta. The magnetic octupole and electric hexadecapole hyperfine interaction constants were calculated by an extended version~\cite{JGLiunpublished} of the HFS92 code~\cite{HFS92} based on the GRASP package~\cite{FroeseFischer2018b}
Our results, labelled as MCDHF, are presented in Table~\ref{Table:HFS}.  

\subsection{Evaluation of theoretical uncertainties}
\label{Sec:TheoreticalErrors}
We start with a comparison of our computed values for $A$ and $B$ HFS constants with the previously published results and then assess our theoretical accuracy.

Comparing our computed values (see Table~\ref{Table:HFS}) with  theoretical values by \citet{Dzuba2016a}, we observe that our $A$ values match, while there is a roughly $10\%$ discrepancy in values of $B$. Itano (cited in Ref.~\cite{Petrasiunas2012}) has previously computed the $A$ and $B$ constants for the $4f^{13}6s^2$ ($J=7/2$) state in $^{171}\mathrm{Yb}^+$ and $^{173}\mathrm{Yb}^+$. Itano has also used the MCDHF method, but his results are markedly different from ours. Since there are no details of calculations given in Ref.~\cite{Petrasiunas2012}, it is difficult to assess the reasons for this difference. We, however, point out that our MCDHF results are in a better agreement with experimental values. 
For example, the deviation is about 20\% between his result and the experimental value $A(^{171}\textrm{Yb}) = 905$~MHz~\cite{Taylor1999}. Multiplying our $A$ constant for $^{173}$Yb by the ratio $\mu(^{171}\textrm{Yb})I(^{173}\textrm{Yb})/\mu(^{173}\textrm{Yb})I(^{171}\textrm{Yb})$, we obtained $A=882$~MHz for $^{171}$Yb, which differs from the measurement~\cite{Taylor1999} by only 3\%. 

Based on these comparisons we conservatively estimate  the uncertainty of our MCDHF calculations to be $\sim 10\%$ for the magnetic dipole and electric quadrupole hyperfine interaction constants. This  estimate is also consistent with that of  Ref.~\cite{Dzuba2016a}, where they  claimed a similar 10\% theoretical uncertainty for these two constants using a different computational method.
We assign a  10\% theoretical uncertainty to the $D$ constant due to its stable convergence trend with the increasing size of the virtual orbital set. However, it is difficult to evaluate the theoretical uncertainty for the $C/\Omega$ constant since it strongly depends on the computational model, as discussed below. We are, however,  confident in the sign and order of magnitude of this octupole constant.

The magnetic octupole HFS constant has proven to be sensitive to the electron correlations, as they flip the sign of the DHF result. 
{We systematically investigated the  dependence of the calculated $C/\Omega$ values on the size of  computational model space, see Table~\ref{CV_effects}. For example, in this table, the results in the ``no opened subshells" row demonstrate the effect of correlation between electrons in the valence subshells. Because the octupole coupling operator has high multiplicity (tensor of rank 3) and we are interested in the  properties of  the $l=3$ $f$-state hole, we include  up  to $l=6$ virtual orbitals in each layer.
The results including the valence-valence correlation show a good convergence pattern (first row of Table~\ref{CV_effects}). However, the convergence pattern worsens  when we start including core-valence correlations by opening core subshells (columns of Table~\ref{CV_effects}). While the results show some degree of convergence, results from an even larger model space would have been more conclusive.   Unfortunately, the largest computation model that we employed already pushes the limits of computational power  at our disposal. Considering the convergence trends of Table~\ref{CV_effects}, we believe that  the sign and the order of magnitude of the computed $C/\Omega$  constant would not change with increasing model space. 
We carried out additional convergence tests that support this conclusion. For example, trends in Table~\ref{CV_effects} indicate that opening the $3p$ and $4p$ subshells  substantially modify the result; so it is plausible that opening the subshell of the same angular momentum, $2p$ subshell, might modify the result further. To test this hypothesis, we  opened the $2p$ subshell for a small model space and found this effect to be negligible.
We take the result obtained with the largest model space as our final value, $C/\Omega = -6 \times 10^{-4} \, \mathrm{MHz/(b\times{\mu_N}})$. 
 }

\begin{table*}[!ht]
\caption{\label{CV_effects} Values of $C/\Omega$ (in units of $\mathrm{kHz/(b\times{\mu_N}})$ ) as a function of the MCDHF computational model space. 
The columns present the trend with opening successively  deeper core subshells: the first row has no core subshells opened, while the last row lists results with the $3s3p3d4s4p4d5p5s$  core subshells opened. 
 The rows compile values obtained by increasing numbers of virtual orbitals. $N^\mathrm{th}$ layer includes $N$ virtual orbitals for  each of the $s, p, d ,f, g, h, i$ angular symmetries. For example, the 1st layer  includes one virtual orbitals for each  $l \in [0,6]$, i.e. 7 orbitals in total. The value marked in bold was obtained with the largest model space.
}
\begin{ruledtabular}
\begin{tabular}{cdddddd}
 &    
\multicolumn{1}{c}{1st layer }&
\multicolumn{1}{c}{ 2nd layer}    &    
\multicolumn{1}{c}{3rd layer }     &    
\multicolumn{1}{c}{4th layer }       &      
\multicolumn{1}{c}{5th layer}        \\
\hline
no opened subshells                
        &     0.578   &        0.630       &     0.649        &     0.652         &       0.652          \\
 $~5s$  &     1.692   &        2.857       &     2.625        &     2.708         &       2.646          \\
 $+5p$  &     2.453   &        3.633       &     3.360        &     3.439         &       3.366          \\
 $+4d$  &     1.986   &        3.072       &     2.778        &     2.822         &       2.749          \\
 $+4p$  &     2.014   &        2.823       &     2.148        &     1.904         &       1.685          \\
 $+4s$  &     2.239   &        3.126       &     2.380        &     2.058         &       1.792          \\
 $+3d$  &     2.131   &        2.956       &     2.173        &     1.822         &       1.541          \\
 $+3p$  &     2.007   &        2.208       &     0.941        &     0.276         &      -0.198      \\
 $+3s$  &     1.932   &        2.037       &     0.688        &     -0.040        &      \bf{-0}.\bf{558}      \\  
 \end{tabular}
\end{ruledtabular}
\end{table*}


As to the second-order corrections $\eta_X$, these are proportional to various products of electronic matrix elements of magnetic-dipole and electric-quadrupole hyperfine interactions. Based on our accuracy estimates for $A$ and $B$,
we conservatively assign $\sim 10\%$ theoretical uncertainty to such matrix elements. Thereby, we expect a $\sim 10\%$ theoretical uncertainty in the second-order HFS constants.
In addition, the second-corrections contain summation over intermediate states; in our calculations we truncated the entire sum to a single contribution from the lowest-energy $F_{5/2}$ state.
We examined contributions from other 12 lowest-energy intermediate states and found that 
$\eta_{\mu\mu}$, $\eta_{\mu{Q}}$, $\eta_{QQ}$ are modified by less than 4\%, 8\%, and 20\%, respectively. Thus the overall theoretical  uncertainty in second-order corrections is in the order of 10\%.


\section{Projected Experimental Accuracy}
\label{sec:Experimental}

\subsection{Experimental Procedure}
The measurement of the hyperfine intervals $\Delta W_i$ of ${}^{173}\mathrm{Yb}^+({}^2\!\!\;F_{7/2}^o)$ can be accomplished via microwave Ramsey spectroscopy on a single trapped ion.  A pure state can be prepared by beginning with optical pumping on the  narrow-band (E2) ${}^2\!\:D_{5/2} \!\leftarrow \! {}^2\!\:S_{1/2} $ transition at $411 \mbox{ nm}$, which will spontaneously decay mainly to ${}^2\!F_{7/2}^o$ via the allowed E1 transition at $\lambda \! = \! 3.4\mbox{ }\mu\mbox{m}$.  By restricting the E2 transition to drive only ${}^2\!D_{5/2}(F\!=\!0) \!\leftarrow \! {}^2\!S_{1/2}(F\!=\!2) $, the $F\!=\!1$ hyperfine level in ${}^2\!F_{7/2}^o$ will be populated.  Following this optical pumping step, resonant microwaves can be used to drive $2\!\leftarrow \! 1$ at $\approx 1\mbox{ GHz}$, followed by de-shelving of the remaining $F\!=\!1$ population in ${}^2\!F_{7/2}^o$ back to ${}^2\!S_{1/2}$ via the E2 transition ${}^1\![3/2]_{3/2}^o \! \leftarrow \! {}^2\!F_{7/2}^o$ at $\lambda\!=\!760 \mbox{ nm}$.  An ion in the ground state can be distinguished from a ${}^2\!F_{7/2}^o$ ion via the appearance or lack of laser-induced fluorescence on ${}^2\!P_{1/2}^o \leftrightarrow \!{}^2\!S_{1/2}$.  By observing how the microwave resonance frequency depends upon the magnetic field in the trap, the $M_F\!=\!0 \! \leftrightarrow \! 0$ transition can be isolated, permitting preparation of the ${}^2\!F_{7/2}^o(F\! = \!2, M_F\! = \!0)$ single quantum state.  From there, stepwise microwave excitation through the hyperfine structure can be used to complete the spectroscopy.  In all cases, read-out is accomplished by observing whether the $760 \mbox{ nm}$ transition de-shelved the ion back to the ground state manifold.

\subsection{Precision}
Since the lifetimes of the states in ${}^{173}\mathrm{Yb}^+(^2\!F_{7/2}^o)$ are all expected to on the order of 1 day or longer \cite{Dzuba2016a}, the achievable precision of these measurements is likely to be limited by practical considerations (as opposed to $T_1$).  In particular, since a small magnetic field will be used to isolate the $M_F\!=\!0 \! \leftrightarrow \!0$ transitions, second-order Zeeman shifts of the clock states can lead to decoherence.  Based on the experimenally determined coherence time of the Zeeman-sensitive hyperfine transitions in the ground state of ${}^{171}\mathrm{Yb}^+$ that we have achieved, we anticipate that $1\mbox{ Hz}$ precision can be obtained by keeping the effective magnetic sensitivity of the ``clock transitions'' ($M_F\!=\!0 \! \leftrightarrow \!0$) in $\mathrm{Yb}^+({}^2F_{7/2}^o)$ below $10^{-3} \mu_\mathrm{B}$.  The offset field required to accomplish this will depend upon how close the zero-field hyperfine states with $\Delta F \! = \! \pm 1$ are to degeneracy.  Assuming there is a pair with significantly smaller zero-field splitting $\Delta W$ than the rest, the effective magnetic moment associated with an offset field $B_\mathrm{o}$ scales as $\mu_\mathrm{eff} \sim B_\mathrm{o}\mu_\mathrm{B}^2/\Delta W$.  Barring any ``accidental'' near-degeneracies ($\Delta W\! < \!10\mbox{ MHz}$), a precision of $1 \mbox{ Hz}$ should be achievable with our current level of magnetic field control.

\subsection{Accuracy}

The potential systematic effects that are expected for this system can be divided into those that will be common to measurements of ground state splittings, and those that are unique to the ${}^2\!\!\;F_{7/2}^o$ state. The former group includes the nonlinear Zeeman shifts from static magnetic fields, differential Stark shifts from the trap fields, blackbody and time-dilation shifts, off-resonant shifts of the levels being measured due to the microwave probe field, and hyperfine-induced third-order corrections \cite{Safronova2010b}.  Since ground-state splittings have been measured below the target precision of $1\mbox{ Hz}$ for many years \cite{Fisk1997,Werth1987}, the techniques to avoid effects such as these have already been demonstrated and are expected to be sufficient for reaching the comparatively modest target accuracy of $1\mbox{ Hz}$.  In particular, taking the expected zero-field splittings from the coefficients in Table \ref{Table:HFS} suggests that the largest second- and fourth-order Zeeman shifts will be on the $F =4 \!\leftrightarrow \!3$ transition, which will contribute a systematic shift of less than $1\mbox{ Hz}$ at $B_\mathrm{o} =5\mbox{ mG}$.

For systematics that are unique to the ${}^2\!\!\;F_{7/2}^o$ state, the largest is anticipated to be the energy shifts from the electronic electric quadrupole interacting with static electric field gradients in the trap. The diagonal contributions to the shifts are be given by
\begin{equation}
\begin{split}
E^{(e.q.)}_{F,M_F}= -e \sum_\mu T_{2,\mu}(\nabla \mathbf{E}) \,\langle \gamma{IJF}|T_{2,-\mu}(\boldgreek{\Theta})|\gamma IJF\rangle\\
= - e \,T_{2,0}(\nabla \mathbf{E})\,\frac{2(3M_F^2 - F(F+1))}{\sqrt{(2F+3)(2F+2)(2F+1)2F(2F-1)}}\\
\times (-1)^{I+J+F} (2F+1)
\frac{\begin{Bmatrix}
    J & F & I \\
    F & J & 2
    \end{Bmatrix}}{
    \threej{J}{2}{J}{-J}{0}{J}}\Theta(\gamma J) ,
\end{split}\label{Eq:DiagonalQuad}
\end{equation}
where the quadrupole moment has been measured to be $\Theta({}^2\!\!\;F_{7/2}^o) = -0.041(5) a_\mathrm{o}^2$ \cite{Huntemann2012}.  These contribute sub-Hz shifts for an electric field gradient of $1\mbox{ kV}/\mbox{cm}^2$, which is significantly larger than the gradient in our current trap.
There are also potentially, off-resonant shifts due to the Paul trap's radiofrequency drive if pairs of states happen to be split by a frequency near the rf drive, in which case the rf drive frequency may need to be changed.  We are therefore not aware of any barriers to achieving a precision of $1 \, \mathrm{ Hz}$ for this measurement.

\section{Discussion}
\label{sec:Discussion}

Equations~(\ref{eq:DeltaW12}) provide the relationship between the 5 quantities that will be measured experimentally (the $\Delta W^{(1+2)}_F$) and the 7 parameters to be determined, $A$-$D$ and the $\eta_{mn}$. However, since all of the terms included in our model are tensors of rank $k\!\le\!4$, there is a degeneracy in Eqs.~(\ref{eq:DeltaW12}) and a proper linear combination of any four of the measurements can be used to predict the fifth.  While this reduces the number of experimentally determined quantities to $k_\mathrm{max}\!=\!4$, it will provide a test of the model presented above and way to detect and reject systematic effects in the experiment.

Within the 3 second order terms ($\eta_{mn}$), since the energy difference between the ${}^2\!F_{J^\prime\!=\!5/2}^o$ and ${}^2\!F_{J\!=\!7/2}^o$ states is known, if we assume that these are the only terms that contribute, they contain only 2 unknowns: $\mu \langle \gamma J || T^e_1 || \gamma J^\prime \rangle$ and $Q \langle \gamma J || T^e_2 || \gamma J^\prime \rangle$. Further, the coefficient $A$ can be determined from existing experimental data \cite{Fisk1997,Werth1987,Taylor1999},
\begin{equation}\label{eq:Acoeffs}
    A^{(173)}_{^2F_{7/2}^o} = \frac{A^{(173)}_{^2S_{1/2}}\,\, A^{(171)}_{^2F_{7/2}^o}}{A^{(171)}_{^2S_{1/2}}} = -250 \mbox{ MHz}.
\end{equation}
Here, we have extracted $A_{{}^2F_{7/2}^o}^{(171)}$ from the measured energy splitting $\Delta W_3^{(171)}=3.620\mbox{ GHz}$ \cite{Taylor1999} via
\begin{equation}
    A_{{}^2F_{7/2}}^{(171)} = \frac{\Delta W_3^{(171)}}{4} + \frac{1}{144}\left( \frac{\mu^{(171)}}{\mu^{(173)}}\right)^2 \eta_{\mu\mu} \approx \frac{\Delta W_3^{(171)}}{4}
\end{equation}
and therefore neglected the contribution (tens of Hz) of the second order correction to the hyperfine splitting of ${}^{171}\mathrm{Yb}^+({}^2\!F_{7/2}^o)$ since it is not expected to contribute at the current level of experimental precision. This term should of course be included in a full treatment when experimental precision reaches the $100\mbox{ Hz}$ level, and adding it does not increase the number of unknowns in the system of equations (\ref{eq:DeltaW12}).
The two ground state $A$ coefficients in (\ref{eq:Acoeffs}) are known to sub-Hz precision \cite{Fisk1997,Werth1987}, and the limiting measurement is $\Delta W_3^{(171)}$, the ${}^2F_{7/2}^o$ HFS splitting in ${}^{171}\mathrm{Yb}^+$ \cite{Taylor1999}.  Using essentially the same procedure as described below, this splitting in ${}^{171}\mathrm{Yb}^+$ can be measured to the same precision (if not better) than the $\Delta W_i$ in ${}^{173}\mathrm{Yb}^+$.  This leaves Eqs.~(\ref{eq:DeltaW12}) with 5 unknowns ($B$, $C$, $D$, $\mu \langle \gamma J || T^e_1 || \gamma J^\prime \rangle$, and $Q \langle \gamma J || T^e_2 || \gamma J^\prime \rangle$).

Because the experimental uncertainty can reach $\sim 1 \, \mathrm{ Hz}$, we expect that the dominant error in extracting first-order HFS constants is due to theoretical uncertainty in the second-order corrections $\eta_{X}$ (see Sec.~\ref{Sec:TheoreticalErrors}). 
One of the possibilities is to determine the second-order corrections directly from the experimental data, but the system of effectively 4 equations and 5 unknowns here will not allow unambiguous extraction of all 5 unknown parameters.

Instead, we solve Eqs.~(\ref{eq:DeltaW12}) for the first four HFS splittings $\Delta W_{F}$ for the HFS constants, $A$, $B$, $C$, and $D$. Each of the resulting equations contains a contribution from the second-order corrections. In particular,  the induced variation in $D$ is $\delta D \approx 3.4 \times 10^{-4} \, \delta \eta_{QQ}$. As discussed in Sec.~\ref{Sec:HFI-YbII}, the  fiducial value of the hexadecapole moment $\Pi \sim 9 \, \mathrm{b}^2$,
leading, in combination with results in Table~\ref{Table:HFS}, to the expected value of $D\approx 2 \, \mathrm{kHz}$. Since $\eta_{QQ}\approx- 200  \, \mathrm{kHz}$, even a 100\% error in $\eta_{QQ}$ would lead to only 3\% error in the extracted value of $D$.
Estimating the induced uncertainty in $C$ is more involved: $\delta C = -1.6 \times 10^{-3} \, \delta\eta_{\mu Q} 
+ 8.9 \times 10^{-4} \, \delta\eta_{QQ}$. If we assume a 10\% error in both $\eta_{\mu Q}$ and $\eta_{QQ}$ per Sec.~\ref{Sec:TheoreticalErrors}, then the induced uncertainty in $C$ is $30\, \mathrm{Hz}$.     
Meanwhile, the expected values of $C$ depends substantially on the assumed value of the octupole moment $\Omega$. 
If we take $\Omega$ from the spectroscopic determination~\cite{Singh} in neutral Yb, the resulting value of $C \approx 21 \, \mathrm{kHz}$;
the nuclear shell model value of $\Omega$ (see Table~\ref{table:NuclearMoments}) yields $C \approx 90 \, \mathrm{Hz}$, and the more sophisticated nuclear model~\cite{Williams1962} reduces $C$ to $2 \, \mathrm{Hz}$. It is clear that for the latter case the uncertainties in the second order correction would mask the contribution of $C$ to the hyperfine splittings and only an upper limit on $\Omega$ can be placed. In such a scenario, one could still determine $D$ and extract the hexadecapole moment, as the value of $D$ is several orders of magnitude larger than $C$.

Given that the well-controlled electronic structure of the ${}^2\!\!\;F_{7/2}^o$ state of $\mathrm{Yb}^+$ should allow for the extraction of measurable, high-order spectroscopic multipole moments, it is possible that even finer detail may be possible.  While nuclear theory suggests that the magnetic multipole moments may be difficult to discern, the electric moments from deformed cores appear straightforward to measure.  In particular, the radioactive ${}^{169}\mathrm{Yb}$ nuclide has spin $I\!=\!7/2$ and a half-life of $\approx \!32$ days, suggesting that precision spectroscopy of the ${}^2\!\!\;F_{7/2}^o$ state of ${}^{169}\mathrm{Yb}^+$ may reveal signatures of its electric 64-pole moment.  The calculation of more 2nd-order correction terms as well as 3rd-order corrections would be required to extract this moment from the data, but we see no fundamental barriers to future studies along these lines.


\acknowledgements
We would like to thank V.~Dzuba for discussions.
This work was supported in part by the U.S. National Science Foundation (Award Numbers 1912555 and 1912465).
JGL is grateful to the University of Nevada, Reno for hospitality and acknowledged the financial support by the National Natural Science Foundation of China (Grant No.~11874090). 

\appendix
\section{Relation between multi-electron and single-electron matrix elements}\label{app:2nd-quantization}
In this Appendix, we prove Eqs.~(\ref{eq:multi-electron-state}, \ref{eq:relation_multi_hole}).

The operator $T_{k,\mu}^e$ in the second quantized form reads~\cite{lindgren1986atomic}
\begin{eqnarray}
\label{eq:T_in_2nd_quantization}
T_{k,\mu}^e=\sum\limits_{i,j}\normprod{a_i^{\dagger}a_j}\bra{i}t_{k,\mu}^e\ket{j},\,
\end{eqnarray}
where $i$ and $j$ represent either core or virtual orbitals, $\bra{i}t_{k,\mu}^e\ket{j}$ is the matrix element, and $\normprod{a_i^{\dagger}a_j}$ are products of creation and annihilation operators in the normal form. We would like to evaluate 
the expectation value of the operator in Eq.~(\ref{eq:T_in_2nd_quantization}) in the many-body state \ket{\psi_{J,\,M}}, Eq.~(\ref{eq:multi-electron-wf}). The intermediate result for the expectation value can be obtained using the Wick's theorem~\cite{lindgren1986atomic},
\begin{eqnarray}
\label{eq:kronecker_delta}
\langle0_c|a_{h'}^{\dagger}a_{v'}a_{w'}\normprod{a_i^{\dagger}a_j}a_w^{\dagger}a_v^{\dagger}a_h|0_c\rangle &=&\nonumber\\ 
-\delta_{ih}\delta_{h'j}(\delta_{vv'}\delta_{ww'}&-&\delta_{v'w}\delta_{w'v})\,\nonumber\\
+\delta_{jw}\delta_{h'h}(\delta_{vv'}\delta_{iw'}&-&\delta_{iv'}\delta_{w'v})\nonumber\\
-\delta_{jv}\delta_{hh'}(\delta_{v'w}\delta_{w'i}&-&\delta_{iv'}\delta_{ww'})\,,
\end{eqnarray}
where $h(h')$ stands for the $4f$ hole orbital and $v(v')$ and $w(w')$ represent the $6s$ orbitals.

Then we immediately obtain 
\begin{equation}
\label{eq:appendix_multi_to_hole_state}
 \bra {\psi_{J,\,M}} T^e_{k,\mu} \ket{\psi_{J,\,M}}  = -\bra{h}t_{k,\mu}^e \ket{h}\,,
\end{equation}
where $\ket{\psi_{J,M}}$ is the multi-electron state of $^{173}\mathrm{Yb}^{+}$, Eq.~{(\ref{eq:multi-electron-wf})}. The reason that the $6s$ orbitals do not contribute to  Eq.~(\ref{eq:appendix_multi_to_hole_state}) is that the operator is non-scalar and the $6s^2$ shell has zero total angular momentum by construction of the multi-electron state~{(\ref{eq:multi-electron-wf})}. 

In general, Eq.~(\ref{eq:appendix_multi_to_hole_state}) works for any non-scalar one-body operator. If we replace $T_{k,\mu}^{e}$ and $t_{k,\mu}^e$ with the $z$ components of the angular momentum operators $J_z$ and $j_z$ respectively in Eq.~(\ref{eq:appendix_multi_to_hole_state}), we obtain the magnetic quantum number of the hole state, $m_h$ equal to $-M$. 

Then, we rewrite Eq.~(\ref{eq:appendix_multi_to_hole_state}) as follows,
\begin{equation}
\label{eq:app_final_multi_single_mel}
 \bra {\psi_{J,\,M}} T^e_{k,\mu} \ket{\psi_{J,\,M}}  = -\bra{\phi_{J,\,-M}}t_{k,\mu}^e \ket{\phi_{J,\,-M}}\,,
\end{equation}
where $\phi_{J,-M}$ is the orbital of the hole-state electron. This proves Eq.~(\ref{eq:multi-electron-state}) of the main text.



Applying the Wigner-Eckart theorem and setting $\mu=0$ on each side of Eq.~(\ref{eq:app_final_multi_single_mel}), we obtain,
	\begin{eqnarray}
	\label{apd:apd1}
 \bra {\psi_{J,\,M}}T^e_{k,0
	}\ket{\psi_{J,\,M}}=&& \nonumber \\
	(-1)^{J-M}\begin{pmatrix}
	J & k & J \\
	-M & 0 & M
	\end{pmatrix} &\langle \psi_{J}||T^e_{k}||\psi_{J}\rangle& \,,
	\end{eqnarray} 

	\begin{eqnarray}
	\label{apd:apd2}
	     -\bra{\phi_{J,\,-M}}t^e_{k,0}\ket{\phi_{J,\,-M}} 
	 = &&
	 \nonumber\\ -(-1)^{J+M}\begin{pmatrix}
	J & k &J \\
	M & 0 & -M
	\end{pmatrix}& \langle \phi_{J}||t^e_{k}(i)||\phi_{J}\rangle&\,.
	\end{eqnarray} 
	\noindent Since $\begin{pmatrix}
	J & k & J \\
	-M & 0 & M
	\end{pmatrix} =(-1)^{2J+k}\begin{pmatrix}
	J & k & J \\
	M & 0 & -M
	\end{pmatrix}$, 
the reduced matrix elements satisfy the following identity,
	\begin{eqnarray}
	\label{apd:apd3}
	    \bra{ \psi_{J}}|T^e_k|\ket{\psi_{J}} &=&(-1)^{1+2M+2J+k} \bra{\phi_{J}}|t^e_{k}|\ket{\phi_{J}}\,\nonumber\\ &=&(-1)^{k+1}\bra{\phi_{J}}|t^e_{k}|\ket{\phi_{J}}.
	\end{eqnarray}
Eq.~({\ref{apd:apd3}}) suggests that when evaluating the reduced matrix elements of even-$k$ operators with multi-electron states, one needs to add an extra negative sign to the single-electron reduced matrix elements. The sign of odd-$k$  reduced matrix elements is unaffected. This proves Eq.(\ref{eq:relation_multi_hole}) of the main text.

Now we generalize these identities to the off-diagonal reduced matrix elements entering the second-order corrections. As discussed in Sec.~{\ref{Sec:HFI-YbII}}, the dominant intermediate state is the $4f^{13}5s^2\,^2\!F_{5/2}$ state denoted as $\ket{\psi_{J'M'}}$. The many-body state $\ket{\psi_{J'M'}}$ has a similar form as Eq.~(\ref{eq:multi-electron-wf}) but differs in the phase factor, $(-1)^{5/2-M'}$ and the annihilation operator $a_{4f_{5/2,-M'}}$.
It can be shown that the relation in Eq.~(\ref{apd:apd3}) still holds for the reduced matrix element,
\begin{eqnarray}
\label{apd:apd4}
	    \bra{ \psi_{J}}|T^e_k|\ket{\psi_{J'}} =(-1)^{k+1}\bra{\phi_{J}}|t^e_{k}|\ket{\phi_{J'}}.
\end{eqnarray}

\section{Reduced matrix elements of hyperfine interaction}\label{app:matrix-element}
Formally, the one-electron wave function is represented by Dirac bi-spinor
    \begin{equation}
        \ket{nj\kappa m}=\begin{pmatrix}
        iP_{n\kappa}(r)\Omega_{\kappa,m}(\hat{r})  \\
        Q_{n\kappa}(r)\Omega_{-\kappa,m}(\hat{r})
        \end{pmatrix},
    \end{equation}
where $P$ and $Q$ are the large and small components of one-electron wave function and $\kappa$ is the relativistic quantum number ($\kappa=\mp{j+\frac{1}{2}}$ for $j=l\pm{\frac{1}{2}}$). The  reduced matrix elements of the electronic part of  hyperfine interaction
are explicitly~\cite{Johnson2007}
\begin{widetext}
\begin{eqnarray}
	\langle n'\kappa'|| t_k^e|| n\kappa \rangle&=&
	\begin{cases}
	-\bra{\kappa^{'}}|C_{k}|\ket{\kappa} \int_{0}^{\infty}\frac{dr}{r^{k+1}}(P_{n'\kappa'}P_{n,\kappa}+Q_{n'\kappa'}Q_{n,\kappa}), & \text{odd k}\,, \\
	\bra{\kappa^{'}}|C_{k}|\ket{-\kappa}\frac{\kappa'+\kappa}{k}\int_{0}^{\infty}\frac{dr}{r^{k+1}}(P_{n'\kappa'}Q_{n,\kappa}+Q_{n'\kappa'}P_{n,\kappa}), & \text{even k},
	\end{cases}\label{eq:reduced-matrix-element}
	\end{eqnarray}
\end{widetext}
where we suppressed $j$ for brevity. The odd and even $k$ sub-cases correspond to electric and magnetic interactions respectively.  

%

\end{document}